# Measurement of magnetic penetration depth and superconducting energy gap in very thin epitaxial NbN films

Anand Kamlapure[1], Mintu Mondal[1*], Madhavi Chand[1], Archana Mishra[1,2], John Jesudasan[1], Vivas Bagwe[1], L. Benfatto[3], Vikram Tripathi[1] and Pratap Raychaudhuri[1†]

[1]*Tata Institute of Fundamental Research, Homi Bhabha Rd., Colaba, Mumbai 400005, India.*
[2]*IIC, Indian Institute of Technology Roorkee, Roorkee, Uttarakhand 247667, India.*
[3] *INFM-CNR Statistical Mechanics and Complexity Cente, University of Rome "La Sapienza", P.le A. Moro 5, 00185 Rome, Italy.*

*Abstract:* We report the evolution of the magnetic penetration depth ($\lambda$) and superconducting energy gap ($\Delta$) in epitaxial NbN films with thickness (*d*) varying between 51-3nm. With decrease in film thickness $T_c$ and $\Delta(0)$ monotonically decreases, whereas $\lambda(0)$ monotonically increases. Our results show that while the values of $\Delta(0)$ and $\lambda(0)$ are well described by Bardeen-Cooper-Schrieffer (BCS) theory, at elevated temperatures, films with *d*≤6.5nm show sudden drop in superfluid density associated with the Kosterlitz-Thouless-Berezinski (KTB) transition. We discuss the implication of these results on the time response of superconducting bolometers made out of ultrathin NbN films.

---

[*] Electronic mail: mondal@tifr.res.in
[†] Electronic mail: pratap@tifr.res.in

In recent years NbN thin films have emerged as a popular material for fabricating sensitive superconducting bolometers[1] capable of detecting up to a single photon[2]. The ease of fabrication of these films through magnetron sputtering and pulsed laser deposition, high transition temperature ($T_c$), the mechanical strength and chemical stability makes NbN a material of choice for this application. However, to obtain the desired sensitivity, superconducting single photon detectors typically use a superconducting layer with thickness, $d$<5nm, lithographically patterned into thin strips. At this thickness, which is smaller than the dirty limit coherence length[3] ($\xi$~5nm), superconducting properties often deviate from their bulk counterparts[4]. In particular, the superconducting energy gap ($\Delta$) and the magnetic penetration depth ($\lambda$) can get strongly affected by the increased role of disorder[5], formation of vortex anti-vortex pairs[6] and thermal phase fluctuations[7]. Since the quantum efficiency and the time response of a superconducting bolometer depends on the free energy difference between the normal state and the superconducting state $\left(F_N - F_S = \frac{1}{2}N(0)\Delta^2\right)$ and the kinetic inductance (per unit length) of the device[8] $\left(L_k = \frac{\mu_0 \lambda^2(T)}{A}, A \text{ is the cross sectional area of the film}\right)$ it is important to understand the evolution of $\Delta$ and $\lambda$ in ultrathin NbN films.

In this letter, we report a direct measurement of $\lambda(T)$ and $\Delta(T)$ as a function of thickness in epitaxial NbN thin films with thickness ($d$) varying between 3-51nm. Epitaxial NbN thin films were grown on (100) oriented MgO substrates by reactive magnetron sputtering, by sputtering a Nb target in Ar/$N_2$ (80:20) gas mixture at a substrate temperature of $600^0$C. Further details of sample growth and characterization have been reported elsewhere[3,9]. The thickness of the films was controlled by controlling the time of deposition keeping all other parameters constant. For films with *d*>20nm, the thickness was measured using a stylus profilometer while for thinner films it was estimated from the time of deposition. λ was measured using a "two coil" mutual inductance technique operating at 60kHz. The main advantage of this technique is that it allows measurement of the absolute value of λ over the entire temperature range up to $T_c$ without any prior assumption about the temperature dependence of λ. In this technique, a 8mm diameter thin superconducting film is sandwitched between a quadrupole primary coil and a dipole secondary coil (figure 1(a)). This technique operates on the principle that the thin superconducting film will partially shield the secondary coil from the magnetic field produced by the primary, the degree of shielding being dependent on λ. The mutual inductance between the primary and the secondary coil is measured as a function of temperature by passing a small a.c. excitation current (1mA) through the primary and measuring the in-phase and out-of-phase induced voltage in the secondary using a lock-in amplifier. λ is determined by evaluating the mutual inductance for different values of λ by numerically solving the Maxwell equations and comparing the measured value with the theoretically calculated value[10]. The quadrupole configuration of the primary coil ensures a fast radial decay of the magnetic field such that edge effects are minimized[11]. The excitation field was kept very low (~7mOe) and the cryostat was shielded from the earth's magnetic field using a mu-metal shield. Δ(T) was measured using a home built low temperature scanning tunneling microscope (STM) on freshly prepared NbN films using Pt-Ir tip. The tunneling conductance $(G(V) = dI/dV)$ versus voltage (V) spectra (averaged over 10 voltage sweeps) were recorded from 4K to $T_c$ on freshly prepared NbN thin films using a lock-in based modulation technique operating at 279Hz and a modulation voltage of 100μV. Resistivity was measured using conventional 4-probe techniques by cutting the samples used for penetration depth measurements into rectangular bars.

Figure 1(b) shows the variation of $T_c$ with film thickness. The real (M') and imaginary part (M'') of the mutual inductance as a function of temperature measured using the penetration depth setup for the thickest and the thinnest film is shown in the inset. All our films show a reasonably narrow peak in M''(T) close to $T_c$. $T_c$, extracted from intersection of two tangents drawn above and below the transition on the M'(T) curve, varies from 15.87K for the 50nm thick film to 9.16K for 3nm thick film.

Figure 2 (a-d) show the STM results for two films with $d$=50nm and $d$=5nm. While the morphology of the thicker films ($d$>10nm) show a granular structure (inset Fig. 2(c)), for $d$<10nm we observe step-like structures (inset Fig. 2(d)) reflecting the step edges on the single crystalline MgO substrate. The surface roughness for all the films is ~1±0.2nm. Figures 2(a-b) show the tunneling spectra at various temperatures. $\Delta$ is extracted by fitting these spectra to the tunneling equation,

$$G(V) = \frac{dI}{dV}\bigg|_V = \frac{d}{dV}\left\{\frac{1}{R_N}\int_{-\infty}^{\infty} N_s(E)\{f(E) - f(E-eV)\}dE\right\}, \quad (1)$$

where, $N_s(E) = \text{Re}\left\{(|E| - i\Gamma)/\left((|E| - i\Gamma)^2 - \Delta^2\right)^{1/2}\right\}$ is the lifetime broadened BCS density of states. While the broadening parameter, $\Gamma(=\hbar/\tau)$, formally incorporated[12] to take into account the lifetime ($\tau$) of the quasiparticle, phenomenologically incorporates all sources of non-thermal broadening in the BCS DOS. We observe that the temperature dependence of $\Delta$(T) (Fig. 2(c)-(d)) closely follows the BCS curve[13] within experimental accuracy.

Figure 3(a) shows the temperature variation of $\lambda^{-2}$(T)$\propto n_s$(T) (where $n_s$ is the superfluid density). $\lambda$ increases from 275nm to 529nm as the thickness decreases from 51nm to 3nm. The value of $\lambda$ for the 51nm film is consistent with earlier measurements which vary between[14] $\lambda \sim 200-400$ nm for NbN films with similar thickness. Since in our films the electronic mean free path[3], $l \ll \xi$, we fit the temperature variation (solid lines) with the dirty limit BCS expression[15],

$$\frac{\lambda^{-2}(T)}{\lambda^{-2}(0)} = \frac{\Delta(T)}{\Delta(0)} \tanh\left(\frac{\Delta(T)}{2k_B T}\right), \quad (2)$$

using $\Delta(0)$ as a fitting parameter. The fit is good[16] barring the two thinnest films where we observe an abrupt drop in $n_s$ associated with the KTB transition close to $T_c$. The best fit values of $\Delta(0)$ plotted (figure 3(b)) along with the ones obtained from tunneling measurements (as a function of $T_c$) agree well with each other. A further consistency check can be obtained by noting the dirty-limit BCS relation[17],

$$\lambda^{-2}(0)_{BCS} = \frac{\pi \mu_0 \Delta(0)}{\hbar \rho_0}, \quad (3)$$

where $\rho_0$ is the normal state resistivity just above $T_c$. The agreement between experimental value of $\lambda(0)$ and $\lambda(0)_{BCS}$ calculated using $\rho_0$ (figure 3(d)) and $\Delta(0)$ (interpolated from figure 3(b)) suggests that the evolution of the ground state properties of thin NbN films can be understood from weakening of the electron-phonon pairing interaction, possibly due to increase in the Coulomb pseudopotential[5] arising from loss of effective screening.

We now focus our attention on the KTB transition observed in $n_s$ in the two thinnest samples where, $d \lesssim \xi$. According to the KTB theory $n_s \propto \lambda_{KTB}^{-2}$ is expected to jump discontinuously to zero when free vortices proliferate in the system[18]. This occurs when $\frac{1}{\lambda^2(T_{KTB})} = \frac{8\pi \mu_0 k_B T_{KTB}}{d \phi_0^2}$ ($\phi_0$ is the flux quantum). However, the deviation of $n_s$ from the BCS curve starts much before than the jump, in contrast to the standard expectation based on the 2D XY[18] model. This apparent discrepancy is however resolved noting that the vortex core energy, $\mu(T)\left(\propto \frac{\hbar^2 n_s(T)d}{4m} = J(T)\right)$ in a real superconductor can deviate significantly from the 2D XY value, namely, $\mu_{XY} = \frac{\pi^2}{2}J$. In particular, when $\mu < \mu_{XY}$ bound vortex-antivortex pairs can substantially renormalize $n_s$ with respect to the BCS value already before than $T_{KTB}$, as has been discussed recently by Benfatto et al.[19] Following this approach [ref. 19] we fit the data (dashed line) treating μ as an adjustable parameter, and we model the intrinsic inhomogeneity of the sample - that accounts for the smearing of the discontinuous jump at $T_{KTB}$ - as a Gaussian distribution of the local superfluid densities (J).[19] The fits give $\mu \approx J$ and $\left\langle (J_0 - \overline{J_0})^2 \right\rangle^{1/2} / \overline{J_0} = \delta / \overline{J_0}$ <0.025 confirming the homogeneous nature of the films.

In conclusion, we have carried out detailed measurement of the evolution Δ and λ with film thickness in very thin epitaxial NbN films. Our results show that these values differ significantly from their bulk counterparts. To highlight the importance of these results, we calculate the kinetic inductance of a 100nm wide stripline grown on a 4nm thick film (λ(0)~500nm) with effective length of 500μm to be 392nH. This is very close to the value[8] (415nH) calculated from the reset time of a single photon detector with similar geometry. This indicates that the time response of state of the art NbN superconducting bolometer is primarily limited by the enhanced penetration depth at small thickness, which should be taken into account in any realistic design of devices based on ultrathin NbN films.

*Acknowledgements:* We would like to thank Subhash Pai for invaluable technical support, Sangita Bose for valuable technical advice on how to reduce r.f. noise in the STM and Sourin Mukhopadhyay for his involvement in the fabrication of STM.

*Figure captions:*

**Figure 1.** (a) Schematic diagram of the coil assembly in our penetration depth measurement setup. The upper (primary) coil has 28 turns with the half closer the film wound in one direction and the farther half wound in the opposite direction. The lower (secondary) has 120 turns wound in the same direction in 4 layers. (b) Variation of $T_c$ with film thickness for epitaxial NbN thin films. The *inset* shows M'(T) and M''(T) for the thickest and the thinnest film.

**Figure 2.** (a-b) Tunneling conductance spectra normalized at 7mV ($G_N$) for the 50nm and 5nm thick films respectively. The successive temperatures are (a) 4.7K, 5.8K, 9.35K, 11K, 11.8K, 12.5K and 13.7K and (b) 5.2K, 6.3K, 7.9K, 8.8K, 9.8K 10.2K and 10.6K respectively. The solid lines show the theoretical fits to the spectra. (c-d) Temperature variation of $\Delta(T)$ (points) along with the expected BCS behavior (lines) for the same two films; the *insets* show the topographic image (106nm by 106nm for (c) and 265nm by 265nm for (d)) at 5K of the surface.

**Figure 3.** (a) Temperature variation of $\lambda^{-2}(T)$ for NbN films with different thickness; (b) $\Delta(0)$ extracted from tunneling measurements (solid triangle) and penetration depth (hollow triangle) as a function of $T_c$; (c) $\lambda(0)$ (solid square) and $\lambda(0)_{BCS}$ (solid circle) as a function of $T_c$; (d) $\rho_0$ as a function of $T_c$.

**Figure 4.** Temperature variation of $\lambda^{-2}$ close to $T_c$ for 6.5nm and 3nm films. The dashed lines show the theoretical fit to the data. The best fit parameters are $\mu \approx \overline{J}$, $\delta = 0.025\overline{J_0}$ for the 3nm sample and $\mu \approx 0.785J$, $\delta = 0.009\overline{J_0}$ for the 6.5nm sample. Intersection with the solid line is where the universal KTB, smeared out by inhomogeneities is expected: $\lambda^{-2}(T) = 8\pi\mu_0 k_B T / d\phi_0^2$.

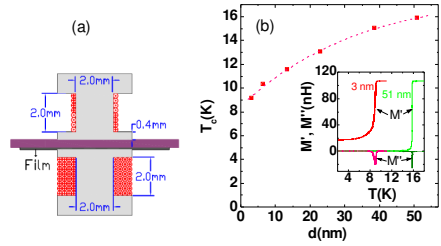

**Figure 1.** (a) Schematic diagram of the coil assembly in our penetration depth measurement setup. The upper (primary) coil has 28 turns with the half closer the film wound in one direction and the farther half wound in the opposite direction. The lower (secondary) has 120 turns wound in the same direction in 4 layers. (b) Variation of $T_c$ with film thickness for epitaxial NbN thin films. The *inset* shows M'(T) and M''(T) for the thickest and the thinnest film.

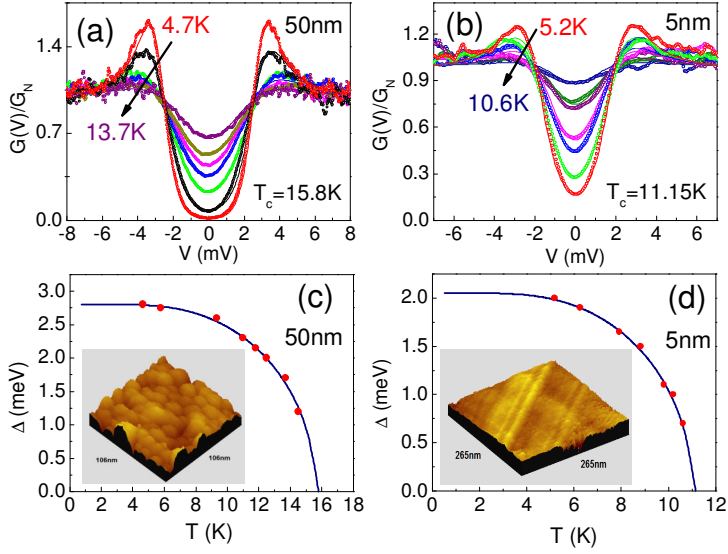

**Figure 2.** (a-b) Tunneling conductance spectra normalized at 7mV ($G_N$) for the 50nm and 5nm thick films respectively. The successive temperatures are (a) 4.7K, 5.8K, 9.35K, 11K, 11.8K, 12.5K and 13.7K and (b) 5.2K, 6.3K, 7.9K, 8.8K, 9.8K 10.2K and 10.6K respectively. The solid lines show the theoretical fits to the spectra. (c-d) Temperature variation of $\Delta(T)$ (points) along with the expected BCS behavior (lines) for the same two films; the *insets* show the topographic image (106nm by 106nm for (c) and 265nm by 265nm for (d)) at 5K of the surface.

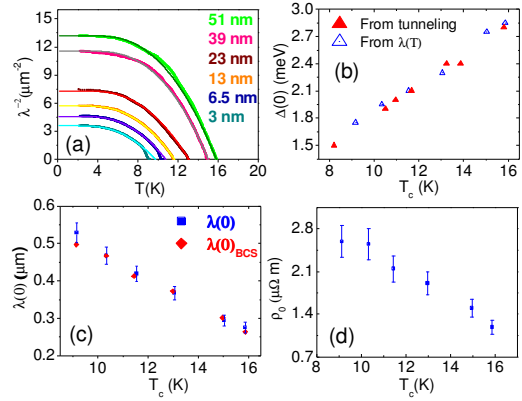

**Figure 3.** (a) Temperature variation of $\lambda^{-2}(T)$ for NbN films with different thickness; (b) $\Delta(0)$ extracted from tunneling measurements (solid triangle) and penetration depth (hollow triangle) as a function of $T_c$; (c) $\lambda(0)$ (solid square) and $\lambda(0)_{BCS}$ (solid circle) as a function of $T_c$; (d) $\rho_0$ as a function of $T_c$.

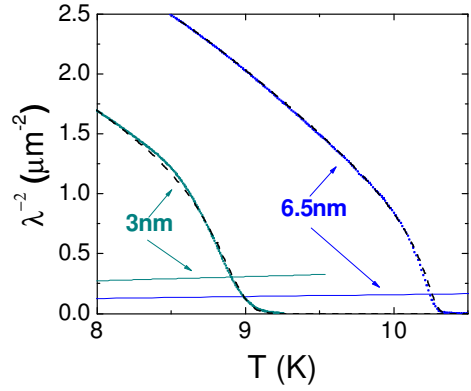

**Figure 4.** Temperature variation of $\lambda^{-2}$ close to $T_c$ for 6.5nm and 3nm films. The dashed lines show the theoretical fit to the data. The best fit parameters are $\mu \approx J$, $\delta = 0.025\overline{J_0}$ for the 3nm sample and $\mu \approx 0.785J$, $\delta = 0.009\overline{J_0}$ for the 6.5nm sample. Intersection with the solid line is where the universal KTB, smeared out by inhomogeneities is expected: $\lambda^{-2}(T) = 8\pi\mu_0 k_B T / d\phi_0^2$.

**References:**

[1] P. Khosropanah, J. R. Gao, W. M. Laauwen, M. Hajenius, and T. M. Klapwijk, Appl. Phys. Lett. **91,** 221111 (2007); A. D. Semenov, H. Richter, H.-W. Hübers, K. S. Il'in, and M. Siegel, Appl. Phys. Lett. **87,** 173508 (2005); J. J. A. Baselmans , A. Baryshev and S. F. Reker, M. Hajenius, J. R. Gao and T. M. Klapwijk, Appl. Phys. Lett. **86,** 163503 (2005).

[2] M. Tarkhov, J. Claudon, J. Ph. Poizat, A. Korneev, A. Divochiy, O. Minaeva, V. Seleznev, N. Kaurova, B. Voronov, A. V. Semenov and G. Gol'tsman, Appl. Phys. Lett. **92,** 241112 (2008); Shigehito Miki, Mikio Fujiwara, Masahide Sasaki, Burm Baek, Aaron J. Miller, Robert H. Hadfield, Sae Woo Nam, and Zhen Wang, Appl. Phys. Lett. **92,** 061116 (2008); S. N. Dorenbos, E. M. Reiger, U. Perinetti, V. Zwiller, T. Zijlstra, and T. M. Klapwijk, Appl. Phys. Lett. **93,** 131101 (2008); M. Ejrnaes, R. Cristiano, O. Quaranta, S. Pagano, A. Gaggero, F. Mattioli, R. Leoni, B. Voronov, and G. Gol'tsman, Appl. Phys. Lett. **91,** 262509 (2007).

[3] S. P. Chockalingam, Madhavi Chand, John Jesudasan, Vikram Tripathi, and Pratap Raychaudhuri, Phys. Rev. B **77**, 214503 (2008).

[4] Z. Wang, A. Kawakami, Y. Uzawa and B. Komiyama, J. Appl. Phys. **79,** 7837 (1996); A. Semenov et al., Phys. Rev. B **80,** 054510 (2009).

[5] A. M. Finkel'stein, Physica B **197,** 636 (1994).

[6] A. F. Hebard and A. T. Fiory, Phys. Rev. Lett. **44,** 291 (1980); S. J. Turneaure, T. R. Lemberger, and J. M. Graybeal, Phys. Rev. B 63, 174505 (2001).

[7] S. J. Turneaure, T. R. Lemberger, and J. M. Graybeal, Phys. Rev. Lett. **84**, 987 (2000).

[8] A. J. Kerman, E. A. Dauler, W. E. Keicher, J. K. W. Yang, K. K. Berggren, G. Gol'tsman and B. Voronov, Appl. Phys. Lett. **88,** 111116 (2006).

[9] Madhavi Chand, Archana Mishra, Y. M. Xiong, Anand Kamlapure, S. P. Chockalingam, John Jesudasan, Vivas Bagwe, Mintu Mondal, P. W. Adams, Vikram Tripathi, and Pratap Raychaudhuri, Phys. Rev. B 80, 134514 (2009); S. P. Chockalingam, M. Chand, J. Jesudasan, V. Tripathi, and P. Raychaudhuri, J. Phys.: Conf. Ser. **150,** 052035 (2009).

[10] S. J. Turneaure, E. R. Ulm, and T. R. Lemberger, J. Appl. Phys. **79,**4221 (1996); S. J. Turneaure, A. A. Pesetski, and T. R. Lemberger, J. Appl. Phys. **83,** 4334 (1998).

[11] For our coil configuration, the induced supercurrent at the edge of an 8mm diameter film is two orders of magnitude smaller than the peak supercurrent density induced in the film.

[12] R. C. Dynes, V. Narayanamurti, and J. P. Garno, Phys. Rev. Lett. **41,** 1509 (1978).

[13] While NbN is in the strong coupling limit with $2\Delta/k_BT_c$~4.2, we do not observe a significant deviation in $\Delta(T)$ from weak coupling BCS behavior.

[14] S. Kubo, M. Asahi, M. Hikita and M. Igarashi, Appl. Phys. Lett. **44**, 258 (1984); J-C. Villegier, N. Hadacek, S. Monso, B. Delaet, A. Roussy, P. Tebvre, G. Lamura and J-Y. Laval, IEEE Transact. Appl. Superconductivity **11,** 68 (2001); D. E. Oates, A. C. Anderson, C. C. Chin, J. S. Derov, G. Dresselhaus and M. S. Dresselhaus, Phys. Rev. B **43,** 7655 (1991); R. Hu, G. L. Kerber, J. Luine, E. Ladizinski, J. Bulman, Transact. Appl. Superconductivity **13,** 3288 (2003); B. Komiyama, Z. Wang and M. Tonouchi, Appl. Phys. Lett. **68,** 562 (1996).

[15] M. Tinkham, *Introduction to Superconductivity* (McGraw-Hill International Edition, Singapore, 1996).

[16] The situation should be contrasted with granular samples where long wavelength phase fluctuations can give rise to linear decrease of superfluid density at low temperatures, e.g. G. Lamura, J.-C. Villégier, A. Gauzzi, J. Le Cochec, J.-Y. Laval, B. Plaçais, N. Hadacek and J. Bok, Phys. Rev. B **65,** 104507 (2002).

[17] T. P. Orlando, E. J. McNiff, S. Foner, and M. R. Beasley, Phys. Rev. B **19**, 4545 (1979).

[18] D. R. Nelson and J. M. Kosterlitz, Phys. Rev. Lett. **39,** 1201 (1977).

[19] L. Benfatto, C. Castelani and T. Giamarchi, Phys. Rev. Lett. **98,** 117008 (2007); L. Benfatto, C. Castelani and T. Giamarchi, Phys. Rev. B **77**, 100506(R) (2008).
14